\documentclass[conference]{IEEEtran}
\IEEEoverridecommandlockouts

\usepackage{cite}
\usepackage{amsmath,amssymb,amsfonts}
\usepackage{algorithmic}
\usepackage{graphicx}
\usepackage{textcomp}
\usepackage{xcolor}
\usepackage{epstopdf}
\usepackage{booktabs} 
\usepackage[hidelinks]{hyperref}

\def\BibTeX{{\rm B\kern-.05em{\sc i\kern-.025em b}\kern-.08em
    T\kern-.1667em\lower.7ex\hbox{E}\kern-.125emX}}
\begin{document}

\title{SmartCourse: A Contextual AI-Powered Course Advising System for Undergraduates\\
\thanks{ * Corresponding Author}
}

\author{\IEEEauthorblockN{1\textsuperscript{st} Yixuan Mi*}
\IEEEauthorblockA{\textit{Department of Computer Science,} \\
\textit{Wenzhou-Kean University,}\\
Wenzhou, China \\
\textit{Department of Computer Science,} \\
\textit{Kean University,}\\
Union, NJ, United States\\
miy@kean.edu}
\and
\IEEEauthorblockN{2\textsuperscript{nd} Yiduo Yu}
\IEEEauthorblockA{\textit{Department of Computer Science,} \\
\textit{Wenzhou-Kean University,}\\
Wenzhou, China \\
\textit{Department of Computer Science,} \\
\textit{Kean University,}\\
Union, NJ, United States\\
yuyid@kean.edu}
\and
\IEEEauthorblockN{3\textsuperscript{rd} Yiyi Zhao}
\IEEEauthorblockA{\textit{Department of Computer Science,} \\
\textit{Wenzhou-Kean University,}\\
Wenzhou, China \\
\textit{Department of Computer Science,} \\
\textit{Kean University,}\\
Union, NJ, United States\\
zhaoyiy@kean.edu}
}

\maketitle

\begin{abstract}
We present SmartCourse, an integrated course management and AI-driven advising system for undergraduate students (specifically tailored to the Computer Science (CPS) major). SmartCourse addresses the limitations of traditional advising tools by integrating transcript and plan information for student-specific context. The system combines a command-line interface (CLI) and a Gradio web GUI for instructors and students, manages user accounts, course enrollment, grading, and four-year degree plans, and integrates a locally hosted large language model (via Ollama) for personalized course recommendations. It leverages transcript and major plan to offer contextual advice (e.g., prioritizing requirements or retakes). We evaluated the system on 25 representative advising queries and introduced custom metrics—PlanScore, PersonalScore, Lift, and Recall—to assess recommendation quality across different context conditions. Experiments show that using full context yields substantially more relevant recommendations than context-omitted modes, confirming the necessity of transcript and plan information for personalized academic advising. SmartCourse thus demonstrates how transcript-aware AI can enhance academic planning. The source code is publicly available at \href{https://github.com/EthanYixuanMi/Smartcourse-Contextual-Advising}{this GitHub repository}.

\end{abstract}

\begin{IEEEkeywords}
Academic advising, Course recommendation, Large Language Models (LLMs), Degree planning
\end{IEEEkeywords}

\section{Introduction}
Traditional academic advising tools often provide generic guidance, lacking integration with student-specific data such as transcripts and degree plans\cite{a15}. Existing advising solutions tend to be generic and fail to incorporate detailed student context (such as completed courses or degree requirements)\cite{a16, a17}. In particular, there is a lack of advising systems that directly integrate a student’s own transcript and major plan into the recommendation process. More recently, advances in AI and LLMs suggest new possibilities for personalized advising\cite{a14}. Chatbot-based advisors can offer tailored course selection advice and career guidance\cite{a18}. Our system supports three types of users—students, instructors, and administrators—each interacting through role-specific interfaces. Motivated by these trends, SmartCourse addresses these limitations by embedding personalized advising directly into the academic workflow. 

We investigate how transcript and degree plan impact course recommendation relevance. To this end, we design controlled experiments that systematically omit these inputs and analyze the resulting drop in recommendation quality.

This paper makes the following contributions:
\begin{itemize}
    \item \textbf{Contextual Advising Architecture}: We develop SmartCourse, a modular platform that tightly couples academic operations with AI-driven advising. SmartCourse combines core academic services with AI advising in a unified interface.
    \item \textbf{AI-Powered Recommendation Engine}: SmartCourse integrates a locally hosted LLM (via Ollama) to generate course suggestions. The LLM generates suggestions based on contextual prompts. Automated email notifications can be sent when new recommendations or updates are available. 
    \item \textbf{Experimental Evaluation with Context Ablation}: We formulate a set of 25 realistic advising queries (e.g. elective choices for AI specialization, courses for a cyber security graduate track, GPA-improvement strategies) and evaluate SmartCourse under four context conditions (full context, no transcript, no plan, question-only). We introduce relevance metrics – PlanScore, PersonalScore, Lift, and Recall – to quantify how well the recommendations align with the student’s outstanding degree requirements, as well as latency in seconds.
    \item \textbf{Results and Insights}: Incorporating transcript and plan context leads to consistently higher recommendation quality, while omitting them harms performance. These findings highlight the importance of contextual information in personalized advising, showing clear improvement over generic recommendations (see Table~\ref{tab:results} for a summary of evaluation scores).
\end{itemize}

The remainder of the paper reviews related work (Section~\ref{sec:related-work}), describes the system design (Sections~\ref{sec:system-overview}–\ref{sec:implementation}), presents the experimental setup and results (Sections~\ref{sec:experiments}–\ref{sec:results}), and discusses key findings, limitations, and future directions (Section~\ref{sec:discussion-limitations-and-future-work}).

\section{Related Work}
\label{sec:related-work}
Course recommendation in education has been studied through collaborative filtering, matrix factorization, and content-based methods\cite{a1, a4, a5}. These techniques often rely on preference data or grade history but typically ignore structured curricular requirements\cite{a7}. Meanwhile, academic dashboards and degree audit tools help students monitor progress and register for courses, but rarely offer personalized or adaptive suggestions. SmartCourse addresses this gap by aligning recommendations with formal degree plans and transcript records.

More recently, LLMs have been explored for educational use cases\cite{a10}, such as chatbot-based tutoring, FAQ answering, or summarizing curriculum content\cite{a3}. While LLMs can offer natural-language interactions, they often lack access to structured academic records, leading to generic or hallucinated advice\cite{a11}.

SmartCourse builds upon these strands by embedding a locally hosted LLM into an end-to-end course management system\cite{a12, a13}. Unlike prior work, it fuses transcript, degree plan, and user queries into contextual prompts, enabling personalized and curriculum-aligned advising within a unified platform.

\section{System Overview}
\label{sec:system-overview}
\subsection{Architecture Overview}
\begin{figure}[htbp]  
    \centering
    \includegraphics[width=0.5\textwidth]{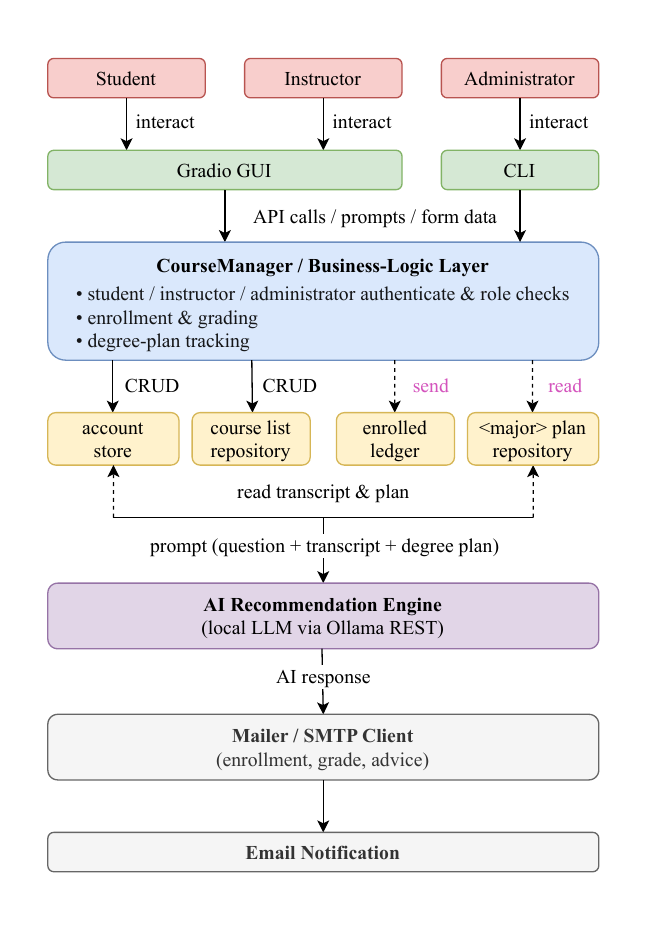}
    \caption{SmartCourse system architecture. Users interact via CLI (administrator) or Gradio GUI (instructor \& student), managed by the  \textit{CourseManager} layer. Data is stored in flat-file repositories, with an LLM providing recommendations and a mailer handling notifications. }
    \label{fig:SmartCourse_arch}
\end{figure}

SmartCourse is organized into several interacting components. It maintains student and instructor accounts (with roles and majors), supports course management (registration, prerequisites, and grades), and enforces a four-year degree plan for the specific major. A central AI Recommendation Engine uses a local LLM to answer student questions based on their academic record. The system provides both a text-based CLI and a user-friendly Gradio web interface for interaction. Fig.~\ref{fig:SmartCourse_arch} illustrates the overall SmartCourse architecture (data inputs, modules, and UI layers).

\subsection{Core Components}
SmartCourse comprises the following building blocks:
\begin{itemize}
    \item \textbf{User Accounts}: Credentials, user roles (student, instructor or administrator), and declared majors are stored in a secure account store with hashed passwords. 
    \item \textbf{Course Enrollment \& Grading}: A course-catalog repository lists all available courses. Students enroll through either interface; instructors record grades, which are written to an enrollment ledger that tracks course codes and grades per student.
    \item \textbf{Degree Plan Management}: For each major (e.g.\ CPS) a standard four-year plan repository defines required courses by year. The system compares this plan against the transcript to monitor progress and identify outstanding requirements.
    \item \textbf{AI Recommendation Engine}: When a student poses a question, SmartCourse constructs a prompt that fuses the question text, current transcript, and four-year plan context. The prompt—structured as transcript, plan, and question—is sent to a locally hosted LLM (specifically \texttt{llama3.1:8b} via the Ollama runtime). The returned answer is parsed and filtered into valid course recommendations (Fig.~\ref{fig:ai_advice}).
    
    \begin{figure}[htbp]  
        \centering
        \includegraphics[width=0.5\textwidth]{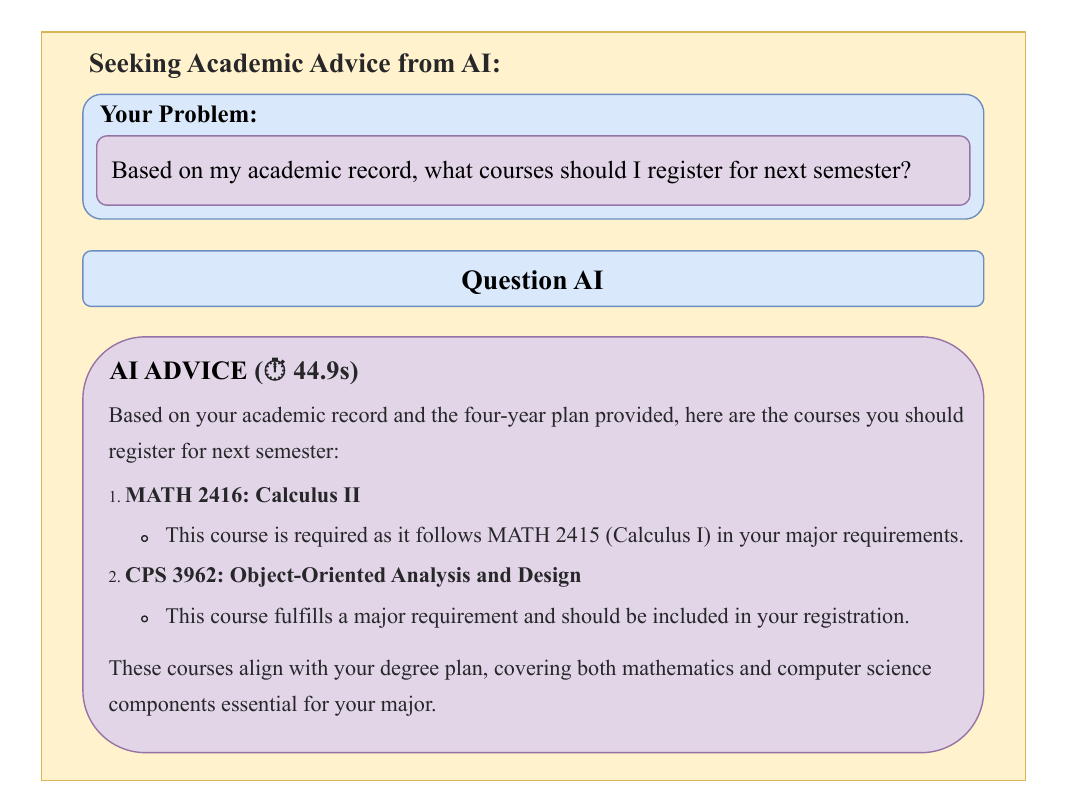}
        \caption{AI-generated advising response based on integrated transcript and degree plan context. \textit{This response was rendered as a schematic interface for clarity and does not depict a live system screen.}}
        \label{fig:ai_advice}
    \end{figure}

    \begin{figure}[htbp]  
        \centering
        \includegraphics[width=0.5\textwidth]{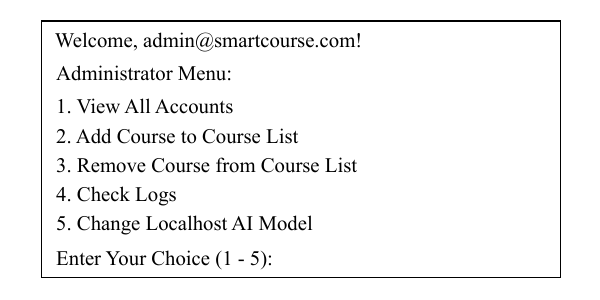}
        \caption{Administrator interface for managing accounts, courses, and AI model configurations via CLI. \textit{Simulated terminal view created for illustrative purposes; not a direct screenshot.}}
        \label{fig:admin_interface}
    \end{figure}
    
    \begin{figure}[htbp]  
        \centering
        \includegraphics[width=0.5\textwidth]{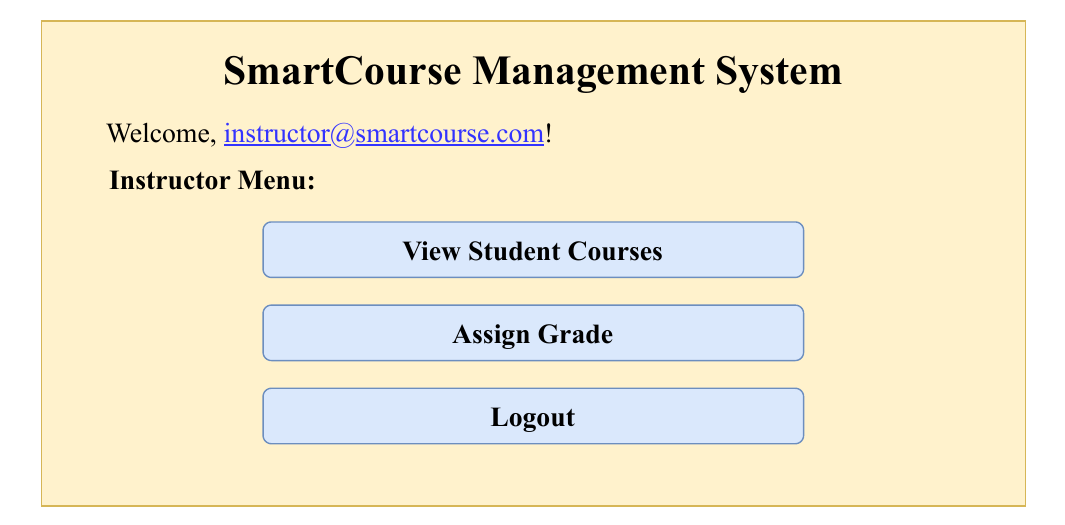}
        \caption{Instructor interface for assigning grades and viewing student course records. \textit{Drawn representation of the interface layout; the actual GUI may differ visually.}}
        \label{fig:instructor_interface}
    \end{figure}

    \begin{figure}[htbp]  
        \centering
        \includegraphics[width=0.5\textwidth]{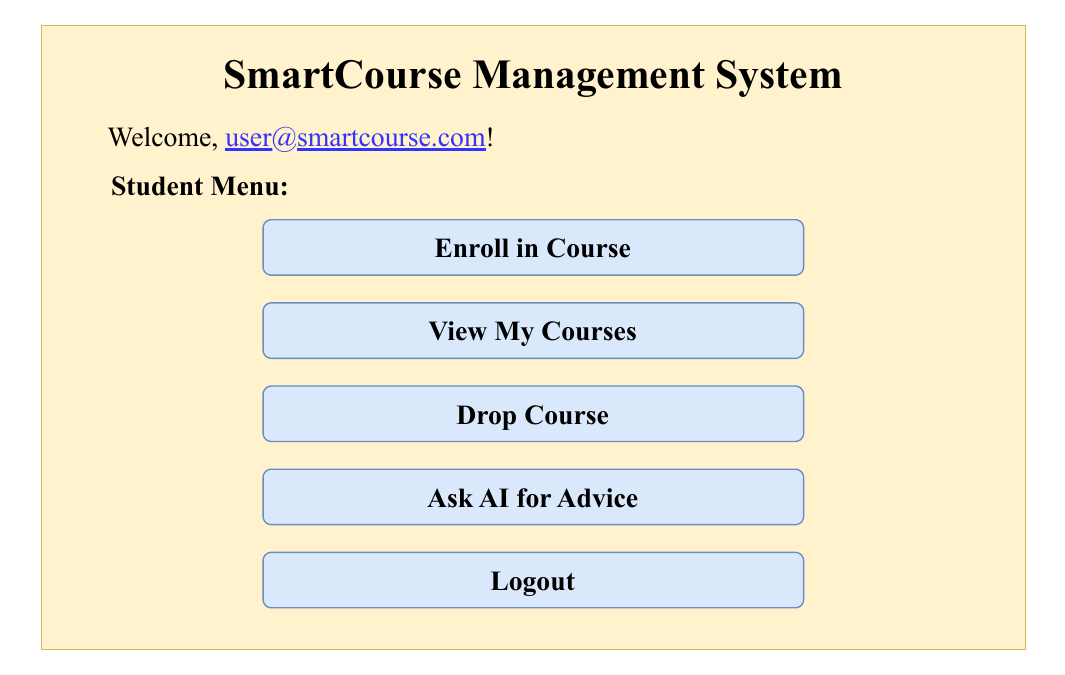}
        \caption{Student interface for registering and dropping courses, viewing progress, and requesting AI-based advising. \textit{This interface depiction is illustrative and not captured from a running system.}}
        \label{fig:student_interface}
    \end{figure}
    \item \textbf{Interfaces}: We illustrate these interfaces below to provide a concrete view of the system’s user experience. Administrators operate through a dedicated CLI to maintain system-level configurations. Their responsibilities include managing the course catalog (adding / removing courses), auditing user accounts, viewing system logs, and switching the local LLM model when needed. Fig.~\ref{fig:admin_interface} shows the administrator CLI menu, which emphasizes control and audit capabilities over advising interactions.
    Instructors (Fig.~\ref{fig:instructor_interface}) and students (Fig.~\ref{fig:student_interface}) interact via a Gradio-based web GUI that supports chat-style queries and form views. For example, a student can ask “Which electives should I take next semester to prepare for a machine-learning track?” and will receive tailored suggestions.
    \item \textbf{Notifications}: A background mailer module dispatches email notifications—e.g.\ new recommendations, enrollment confirmations, or grade postings—to the relevant users through a configurable SMTP gateway.
\end{itemize}

These components work together to deliver an end-to-end advising workflow.  
Fig.~\ref{fig:SmartCourse_arch} illustrates the data flow from user query, through the LLM engine, to the user interface.

\section{Implementation}
\label{sec:implementation}
\subsection{Technology Stack and Data Storage}
SmartCourse is implemented entirely in Python. For ease of deployment and
maintenance, it relies on lightweight \emph{flat-file repositories} rather than
a full database:  
\begin{itemize}
    \item a \emph{secure account store} that keeps hashed user credentials,
          roles, and declared majors;
    \item a \emph{course-catalog repository} that lists every available course
          with its code and title;
    \item a \emph{degree-plan repository} that records the standard four-year
          curriculum for each major; and
    \item an \emph{enrollment ledger} that tracks, for each student, the courses
          taken and the grades awarded.
\end{itemize}

All repositories are plain UTF-8 text files and are read or written through a
thin data-access layer.

\subsection{Functional Modules}
\begin{itemize}
    \item A data module parses files into in-memory structures. The CLI (for admins) and GUI (for students / instructors) authenticate via a secure store.
    \item On student queries, SmartCourse fuses the transcript, plan, and question into a structured prompt, which is sent via Python subprocess to a local Ollama-based LLM.
    \item A lightweight mailer sends notifications such as recommendations or grades.
\end{itemize}

The entire system is parameterized: SMTP credentials, model name, and
repository paths are configurable. Logs are maintained for auditing, and the system supports configurable parameters such as SMTP credentials, model name, and repository paths to ensure deployment flexibility.

\section{Experiments}
\label{sec:experiments}
\subsection{Experimental Setup}
To evaluate the importance of context in course advising, we designed a set of 25 hypothetical student queries and compared SmartCourse’s recommendations under four input conditions. This setup constitutes an ablation study: by selectively removing transcript or degree-plan inputs, we aim to quantify their individual and combined contributions to recommendation quality.
Two illustrative prompts are:
\begin{itemize}
    \item “What elective courses should I choose next semester to strengthen my
          AI foundation, considering the AI courses I have already taken?”
    \item “Which electives would best prepare me for a Ph.D.\ track in Machine
          Learning?”
\end{itemize}

For every prompt we compared four context settings:
\begin{itemize}
    \item \textbf{Full Context} — the LLM receives both the student’s transcript
          and four-year degree plan.
    \item \textbf{No Transcript} — only the degree plan is provided.
    \item \textbf{No Plan} — only the transcript is provided.
    \item \textbf{Question-Only} — the model sees the question text alone.
\end{itemize}

All experiments were conducted using the \texttt{llama3.1:8b} model hosted locally through the Ollama runtime environment, ensuring consistent responses across all context conditions.

\subsection{Dataset}
The study centers on one CPS student profile and its associated academic artifacts:
\begin{itemize}
    \item a \emph{degree-plan repository} listing \textbf{39} required courses;
    \item a \emph{transcript record} showing \textbf{21} of those courses
          already completed, including several low grades (e.g., B– or below), leaving \textbf{18} outstanding;
    \item a \emph{course-catalog repository} containing \textbf{75}
          courses; and
    \item a \emph{query set} of \textbf{25} representative advising questions.
\end{itemize}

Table~\ref{tab:results} summarizes these statistics.  Because no external
ground-truth answers exist, we treat the degree plan as the reference set and
evaluate recommendation quality using the relevance metrics defined in
Section~\ref{sec:evaluation-metrics}, supplemented by manual inspection.

\begin{table*}[t]
\centering
\caption{Average recommendation quality across 25 advising questions under four context settings.}
\label{tab:results}
\begin{tabular}{lcccccc}
\toprule
Mode & \#Rec & PlanScore & PersonalScore & Lift & Recall & Latency (s) \\
\midrule
full          & \textbf{6.56} & 0.53 & \textbf{0.78} & \textbf{0.25} & 0.15 & 47.65 \\
noPlan        & 2.24 & 0.03 & 0.19 & 0.16 & 0.01 & 25.36 \\
noTranscript  & 6.20 & \textbf{0.60} & 0.69 & 0.09 & \textbf{0.17} & 34.34 \\
question      & 0.04 & 0.04 & 0.04 & 0.00 & 0.00 & 21.52 \\
\bottomrule
\end{tabular}
\end{table*}

\subsection{Evaluation Metrics} 
\label{sec:evaluation-metrics}
Let $\mathcal{R}$ be the set of courses recommended by the LLM,  
$\mathcal{P}$ the set of outstanding degree–plan requirements  
(i.e.\ courses in the plan that the student has \emph{not yet} taken), and  
$\mathcal{L}$ the set of courses the student \emph{has} taken but with a low
grade (below~B$-$).  Using these sets we compute:

\begin{itemize}
    \item \textbf{PlanScore} —
          fraction of recommendations that satisfy an unmet plan requirement:
          \[
             \text{PlanScore} \;=\;
             \frac{\lvert \mathcal{R} \cap \mathcal{P} \rvert}{\lvert \mathcal{R} \rvert}.
          \]

    \item \textbf{PersonalScore} —
          fraction of recommendations that either meet an unmet plan
          requirement or suggest retaking a low-grade course:
          \[
             \text{PersonalScore} \;=\;
             \frac{\lvert \mathcal{R} \cap (\mathcal{P} \cup \mathcal{L}) \rvert}
                  {\lvert \mathcal{R} \rvert }.
          \]

    \item \textbf{Lift} — improvement from personalising to low grades:
          \[
             \text{Lift} \;=\; \text{PersonalScore} - \text{PlanScore}.
          \]

    \item \textbf{Recall} — coverage of the student’s remaining plan courses:
          \[
             \text{Recall} \;=\;
             \frac{\lvert \mathcal{R} \cap \mathcal{P} \rvert}{\lvert \mathcal{P} \rvert }.
          \]

    \item \textbf{Latency} — wall-clock time (in seconds) for the LLM
          to produce a complete answer.
\end{itemize}

PlanScore gauges alignment with curriculum requirements; PersonalScore adds
sensitivity to prior performance; Lift isolates the benefit of that
personalization; and Recall measures how comprehensively the remaining plan is
covered.  Latency captures practical responsiveness.  Each of the 25 questions
was evaluated in all four context modes, and metric means with 95\% confidence
intervals were obtained via 10,000 bootstrap iterations. Our metric design follows calls in recommender-systems research to evaluate beyond accuracy alone, incorporating coverage and user-need alignment\cite{a19, a20}.

\section{Results}
\label{sec:results}
SmartCourse successfully generated course recommendations for all queries under the different modes. 

Under the full context setting, the system recommended an average of $\sim$ \textbf{6.6} courses per query. 
The mean \textbf{PlanScore} was about \textbf{0.53}, and the mean \textbf{PersonalScore} was \textbf{0.78}. 
Consequently, the average \textbf{Lift} (Personal – Plan) was about \textbf{0.25}. 
The average \textbf{Recall} was \textbf{0.15}, and the mean latency was about \textbf{48} seconds.

By contrast, omitting context degraded performance significantly. In \textbf{No Plan} mode (transcript only), the mean \textbf{PlanScore} dropped to \textbf{0.03} and \textbf{Recall} to \textbf{0.01}. In other words, very few needed courses were identified without providing the plan. In \textbf{No Transcript} mode (plan only), \textbf{PlanScore} remained relatively high (\textbf{0.60}) because almost all recommendations could come from the plan, but \textbf{Recall} (\textbf{0.17}) was only slightly better than full context. The \textbf{Question-Only} mode (no context) performed worst: \textbf{PlanScore} and \textbf{PersonalScore} were both \textbf{0.04}, with \textbf{Recall} near \textbf{0.00}. In many question-only cases the LLM either returned no course suggestions or unrelated advice.

\begin{figure}[!t]
  \centering
  \includegraphics[width=\linewidth]{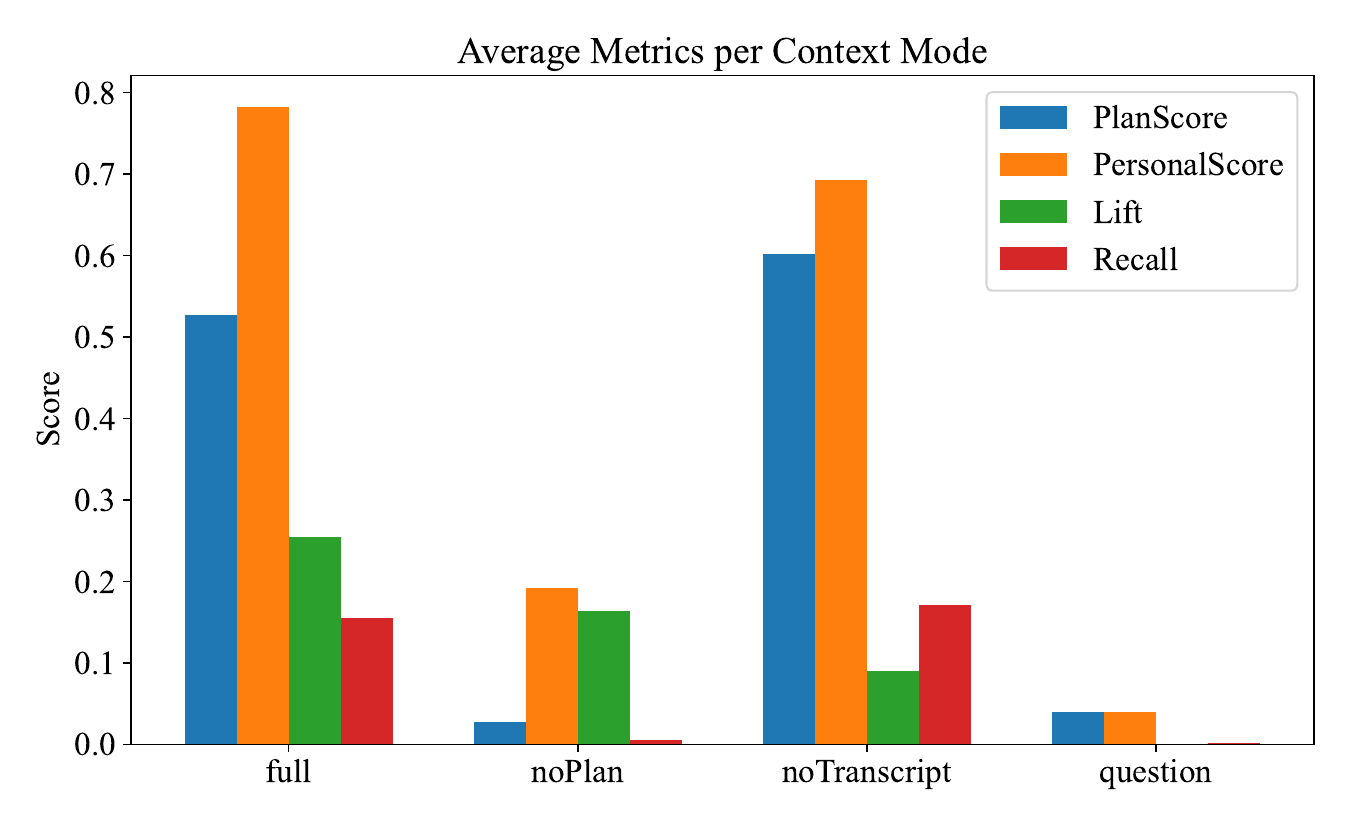}
  \caption{Comparison of average evaluation metrics (PlanScore, PersonalScore, Lift, Recall) across the four context modes. Notably, the full-context mode yields the highest PersonalScore and Lift, while PlanScore is slightly lower than the plan-only mode. This reflects the tradeoff between personalized and strictly curriculum-based recommendations.}
  \label{fig:bar_chart}
\end{figure}

These trends (Fig.~\ref{fig:bar_chart}) illustrate that providing both transcript and plan context is crucial for relevant recommendations. The higher \textbf{PlanScore} and \textbf{Recall} in the full context mode confirm that SmartCourse’s advice aligns well with curriculum requirements. Removing the degree plan leaves the model guessing without structure (hence almost no plan coverage), while removing the transcript (providing only the plan) tended to produce plan-based suggestions but without regard to the student’s actual performance or prerequisites.

Although the \textbf{noTranscript} mode (plan only) yielded slightly higher \textbf{PlanScore} (\textbf{0.60} vs. \textbf{0.53}) and \textbf{Recall} (\textbf{0.17} vs. \textbf{0.15}) than full context, we interpret this result with caution. 
This reflects that, when only the degree plan is available, the model tends to suggest courses directly from the plan, without accounting for the student's progress or past performance. 
In contrast, the full context includes the student's transcript, enabling more personalized suggestions—such as repeating a low-grade course or avoiding completed ones. 
This nuance reduces raw alignment scores like \textbf{PlanScore}, but enhances overall recommendation quality, as captured by the higher \textbf{PersonalScore} (\textbf{0.78}) and \textbf{Lift} (\textbf{0.25}) in full context.

\section{Discussion, Limitations, and Future Work}
\label{sec:discussion-limitations-and-future-work}
\subsection{Discussion}
\label{sec:discussion}
The experimental results highlight the value of SmartCourse’s design. With full context, the system achieved the best overall recommendation performance across key metrics. It consistently suggested appropriate and personalized courses that help the student progress toward graduation. The LLM consistently suggested relevant electives, including upper-level courses and appropriate retakes based on the student’s transcript. It also occasionally proposed retaking a course in which the student earned a low grade, which is a personalized insight a human advisor might provide. In contrast, the No Plan mode essentially lost track of degree requirements, producing mostly irrelevant electives. The near-zero recommendations in the Question-Only mode underline that the LLM needs structured context to answer these academic advising queries meaningfully.

Qualitatively, the advice in full-context mode was coherent. The LLM generally followed logical prerequisites and course progression. We observed, however, occasional hallucinations: suggestions that were not in the official course list or not relevant to the major. For example, in one case the model mentioned a non-CPS elective outside the curriculum. These instances reflect known LLM limitations: without explicit filtering, the model can invent plausible but incorrect items\cite{a22}. In practice, we mitigate this by post-filtering recommended courses against the known course catalog.

\subsection{Current Limitations}
\label{sec:limitations}
Despite encouraging results, several limitations should be acknowledged:
\begin{itemize}
    \item \textbf{Limited Evaluation Scope}: Our experiments used a single student profile and a fixed set of hypothetical questions. Real student needs are far more diverse. The metrics here capture curriculum alignment but not student satisfaction or learning outcomes. More extensive testing with varied majors, true student data, and feedback would be needed.
    \item \textbf{LLM Hallucinations and Bias}: As noted, the LLM can produce incorrect or inappropriate suggestions. We rely on filtering to remove non-existent courses, but subtle biases (e.g., favoring popular electives) may persist. Future work must carefully monitor and correct AI biases.
    \item \textbf{Privacy and Data Security}: SmartCourse processes sensitive academic records. In a deployed environment, strict compliance with FERPA\cite{a23} / GDPR\cite{a24} is essential. (For example, student data must be encrypted and access controlled.) These issues are not fully addressed in the prototype.
    \item \textbf{Incomplete Evaluation Metrics}: Our relevance metrics assume the degree plan is “ground truth.” They do not penalize the system for omitting useful elective suggestions outside the plan. Also, in edge cases where a student has finished nearly all courses, Recall becomes ill-defined. Metrics like precision or qualitative user studies are needed for a fuller evaluation.
    \item \textbf{Latency}: The response time in full-context mode ($\sim$48 seconds) remains the highest among all settings, which may limit real-time usability in practice. While not affecting correctness, it limits usability in interactive advising. Potential optimizations are discussed in Section~\ref{sec:directions}.
\end{itemize}

These limitations point to future research directions. In particular, robustness against AI errors, scalability to multiple programs, and compliance with educational data standards will be critical for real-world use.

\subsection{Directions for Improvement}
\label{sec:directions}
Building upon these limitations, we outline several directions to strengthen SmartCourse’s robustness, personalization, and usability.

\begin{itemize}
    \item \textbf{Broader Major Support}: Currently we use a fixed CPS plan. We aim to generalize SmartCourse to other majors by loading corresponding plan files. This will require curating degree requirements for each program.
    \item \textbf{User Feedback Loop}: Incorporating student or advisor feedback (e.g. a “like / dislike” on recommendations) could refine future suggestions.
    \item \textbf{Rich Contextual Data}: Beyond transcripts, incorporating extracurricular activities or official course descriptions could ground suggestions more effectively.
    \item \textbf{Interface Enhancements}: Developing a full web portal (beyond Gradio) with interactive schedule planners, and real-time notification dashboards, would improve usability.
    \item \textbf{Performance Optimization}: The high latency observed under full context mode could hinder real-time use. Techniques such as prompt caching, model distillation, or asynchronous prefetching could mitigate this\cite{a21}.
\end{itemize}

SmartCourse represents a first step toward AI-augmented academic advising. While the current results validate its core design, transitioning from research prototype to institutional deployment will require empirical evaluation through pilot studies with real students and advisors.

\section{Conclusion}
We have presented SmartCourse, a novel academic advising system that integrates course administration with AI-driven recommendations. By leveraging a student’s transcript and four-year plan, SmartCourse generates contextually relevant course suggestions for a variety of academic queries. Our architecture (Fig.~\ref{fig:SmartCourse_arch}) combines traditional enrollment management with an AI engine, and experimental results confirm that context-aware advising significantly outperforms context-free modes. While challenges remain (LLM reliability, data privacy, scalability), SmartCourse demonstrates the promise of combining institutional data with language models to enhance student guidance. Future work will focus on system refinement, user studies, and broader integration with university information systems. Future extensions will explore integration with student feedback loops and university portals.

\end{document}